\documentclass[%
 aip,
 jmp,%
 amsmath,amssymb,
 reprint,%
]{revtex4-1}

\usepackage{graphicx}% Include figure files
\usepackage{dcolumn}% Align table columns on decimal point
\usepackage{bm}% bold math
\begin{document}

\preprint{AIP/123-QED}

\title{3d-4f spin interaction induced giant magnetocaloric effect in zircon-type DyCrO$_4$ and HoCrO$_4$ compounds}

\author{A. Midya, N. Khan, D. Bhoi and P. Mandal}
\email{prabhat.mandal@saha.ac.in}
\affiliation{Saha Institute of Nuclear Physics, 1/AF Bidhannagar, Calcutta 700 064, India}
\date{\today}

\begin{abstract}
We have investigated the influence of 3d-4f spin interaction on magnetic and magnetocaloric properties of DyCrO$_4$ and HoCrO$_4$ compounds by magnetization and heat capacity measurements. Both the compounds exhibit complicated magnetic properties   and  huge magnetic entropy change around the ferromagnetic transition due to the strong competition between ferromagnetic and antiferromagnetic superexchange interactions. For a field change of 8 T, the maximum values of magnetic entropy change ($\Delta S_{M}^{max}$), adiabatic temperature change ($\Delta T_{ad}$), and refrigerant capacity (RC) reach 29 J kg$^{-1}$ K$^{-1}$, 8 K, and  583 J kg$^{-1}$, respectively for DyCrO$_4$ whereas  the corresponding values for HoCrO$_4$ are 31 J kg$^{-1}$ K$^{-1}$, 12 K, and 622 J kg$^{-1}$. $\Delta S_{M}^{max}$, $\Delta T_{ad}$, and RC are also quite large for a moderate field change. The large values of  magnetocaloric parameters suggest that the zircon-type  DyCrO$_4$ and HoCrO$_4$ could be the potential magnetic refrigerant materials for liquefaction of hydrogen.\\

\end{abstract}


\maketitle
Scientists and engineers are engaged in exploring new refrigeration technology such as  magnetic refrigeration to restrict the use of environmentally harmful substance which is used in vapor compression technology. The refrigeration based on magnetocaloric effect (MCE) is an environment-safe technology\cite{kag,tishin,bfy}. Besides, the magnetic refrigeration unit can be compact so that the  entropy density of magnetic material is larger than that of refrigerant gas and  the efficiency of magnetic refrigeration is higher than  the vapor compression refrigeration \cite{bfy}. The working principle of MCE  describes an isothermal change in the magnetic entropy ($\Delta S_{M}$) or an adiabatic change in the temperature ($\Delta T_{ad}$) when the sample is subjected to a changing magnetic field. As the magnetization changes abruptly and shows strong field dependence near the magnetic phase transition, a large magnetic entropy change is expected to occur close to the transition temperature. Large  MCE in the low-temperature region would be useful for some specific technological applications such as space science, liquefaction of hydrogen in fuel industry while the large MCE close to room temperature can be used for domestic and industrial refrigerant purposes \cite{kag,bfy}. \\

In recent past, several rare-earth transition metal oxides  have been studied due to their interesting physical properties \cite{tk,ima,lee}. The magnetic structure in these systems is composed of  rare-earth and transition metal sublattices.  Generally, the rare earth sublattice (4f) orders antiferromagnetically at low temperature and may show a field-induced  antiferromagnetic (AFM) to  ferromagnetic (FM) transition and large MCE \cite{midya1,jin,shao}. The transition metal sublattice (3d) orders at a relatively high temperature and the nature of magnetic interaction can be FM or AFM depending on crystal geometry \cite{tk,ima}. As the ordering temperatures of two sublattices differ significantly, the transition metal sublattice hardly contributes to  MCE at low temperature. However, $R$CrO$_4$ type compounds ($R$$=$rare earth ions) are exceptional. The neutron diffraction studies and magnetic measurements on  $R$CrO$_4$ show that both the sublattices order collinearly and  simultaneously at a common transition temperature \cite{ylon,pasc}. For smaller ionic radius of $R$ ion, $R$CrO$_4$  usually crystallize in the zircon-type structure with space group $I4_1/amd$ (Ref. \cite{pasc}). The crystal structure of this phase consists of CrO$_4$ tetrahedra and $R$O$_8$ bisdisphenoid polyhedra. Along the two crystallographic axes, $R$O$_8$ units connect with one another by sharing their O-O edges. On the other hand, along the third direction, $R$O$_8$ alternately align with CrO$_4$  units, as a result, the $B$-site CrO$_4$ units are spatially isolated by $R$O$_8$ polyhedra. Because of the presence of two different types of paramagnetic (PM) cations, namely Cr$^{5+}$ and $R^{3+}$,  one expects an interesting magnetic phenomenon due to the magnetic interaction between the 3d and 4f electrons in these $R$CrO$_4$ oxides. As the magnetic entropy  depends on the total angular momentum $J$, the introduction of Cr$^{5+}$ at $X$ site in the isostructural $RX$O$_4$ ($X$$=$P, V, As)  increases the total angular momentum and, therefore, a large entropy change may occur near the  transition temperature of $R$CrO$_4$.\\

In this  work, the field and temperature dependence of magnetic properties of zircon-type compounds DyCrO$_4$ and HoCrO$_4$  have been presented. We have selected these two compounds because Ho$^{3+}$ and Dy$^{3+}$ possess large angular momentum. We observe that both the materials exhibit giant MCE and large adiabatic temperature change and refrigerant capacity (RC) due to the field-induced metamagnetic transition as a result of strong competition between AFM and FM interactions of 3d and 4f spins. The present results also indicate some important differences in the magnetic properties of DyCrO$_4$ and HoCrO$_4$, though both the compounds have almost same magnetic moment. \\

The polycrystalline samples of zircon-type $R$CrO$_4$ ($R$$=$Dy and Ho) have been prepared by standard solid-state reaction method by mixing  the stoichiometric amounts of  $R$(NO$_3$)$_3$:6H$_2$O and Cr(NO$_3$)$_3$:9H$_2$O (Alfa Aesar, purity $>$ 99.99$\%$).  In order to stabilize the unusual Cr$^{5+}$ oxidation state,  the resultant mixture was heated in two steps in flowing  oxygen according to the protocol: 100 $^o$C/h to 200 $^o$C for 1 h and 120 $^o$C/h to 580 $^o$C for 5 h. Finally, the sample was cooled down slowly to room temperature.    No trace of impurity phases were detected within the resolution of powder x-ray diffraction with CuK$_{\alpha}$ radiation (Rigaku TTRAX II).  All the peaks in the diffraction pattern were fitted well to a zircon-type structure of space group $I4_1/amd$ using the Rietveld analysis. The calculated lattice parameters are found to be in good agreement with the earlier report \cite{yl}. We have done thermogravity  analysis  to determine the oxygen content in the sample and energy dispersive x-ray  to check the sample homogeneity. These analyzes show that the oxygen content is close 4 and the sample is homogeneous with its composition close to that of nominal one. Magnetization ($M$) and zero-field heat capacity ($C_p$) measurements were done in a physical properties measurement system (Quantum Design). For the magnetization measurements at a given temperature, the sample was cooled down  from the paramagnetic state to the desired temperature in absence of external magnetic field before collecting the $M$($H$) data. 
\begin{figure}
\begin{center}
\includegraphics[width=.5\textwidth]{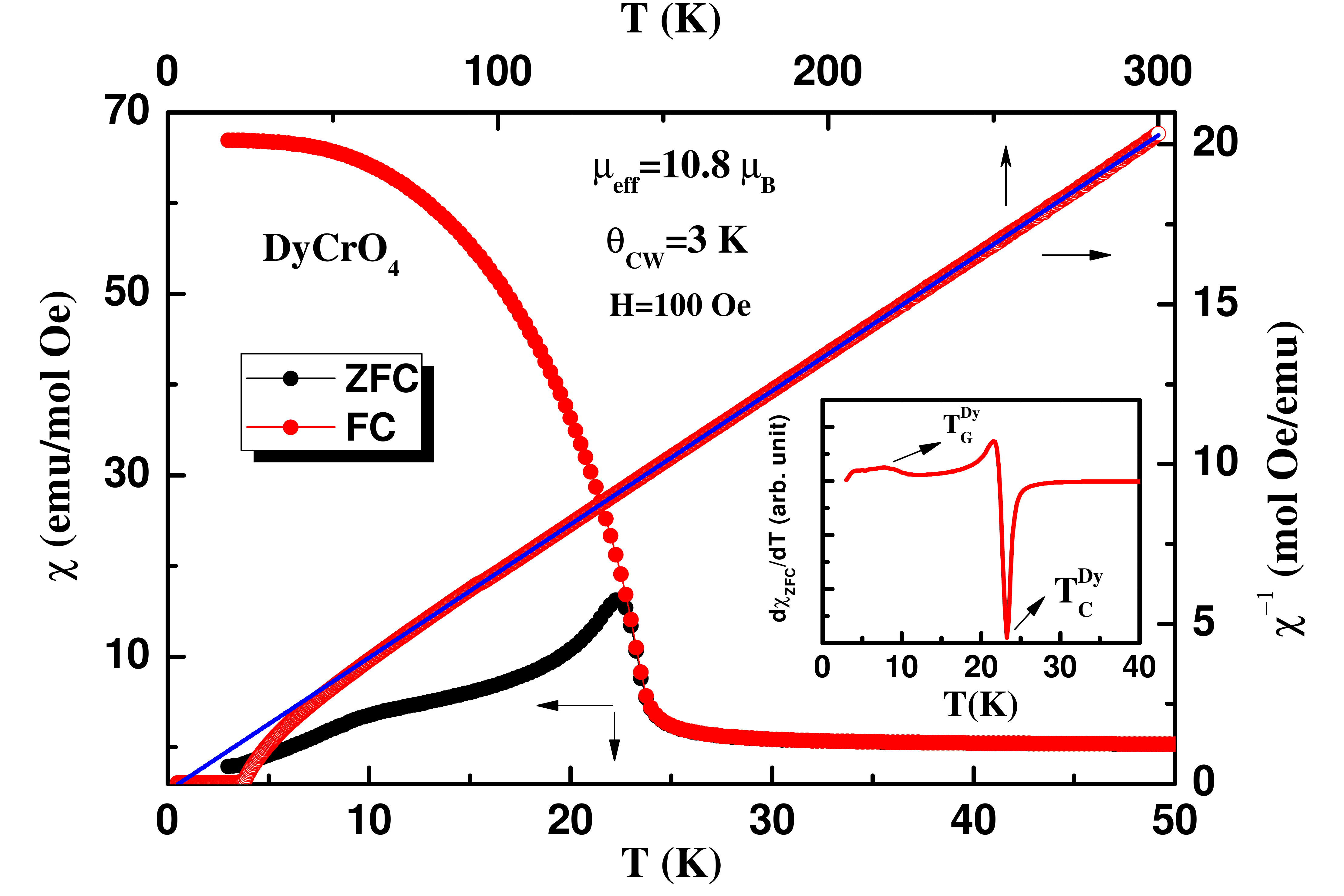}
\caption{ Temperature dependence of FC susceptibility ($\chi_{FC}$) and ZFC susceptibility ($\chi_{ZFC}$) for DyCrO$_4$  at 100 Oe.   The sharp increase in both $\chi_{FC}$ and $\chi_{ZFC}$ below $T_C^{Dy}$$=$23 K is  due to the paramagnetic to ferromagnetic transition. The right axis shows the inverse of FC susceptibility  [$\chi_{FC}$$^{-1}(T)$]  and the corresponding Curie-Weiss fit (solid line). Inset shows the variation of $d\chi_{ZFC}/dT$ with temperature. The weak anomaly at $T_G^{Dy}$$=$8 K is due the spin-glass transition}\label{qmr}
\end{center}
\end{figure}
Figure 1 shows the temperature dependence of zero-field-cooled (ZFC) susceptibility ($\chi_{ZFC}$) and field-cooled (FC) susceptibility ($\chi_{FC}$)  for DyCrO$_4$ measured at a field of 100 Oe. In the PM state, $\chi_{FC}$ and $\chi_{ZFC}$ do not split from each other, and increase monotonically with decreasing $T$. Both $\chi_{FC}$ and $\chi_{ZFC}$ start to increase sharply  due to the PM-FM transition ($T_C^{Dy}$) as $T$ decreases to 23 K. The nature of $T$ dependence of $\chi_{FC}$ and $\chi_{ZFC}$ in the FM state is quite different; $\chi_{FC}$ continue to increase with decreasing $T$ like a Brillouin function as in the case of a typical FM system while $\chi_{ZFC}$ shows an antiferromagnetic-like steep drop slightly below $T_C^{Dy}$. The  decrease in $\chi_{ZFC}$ below $T_C$ is  common in rare-earth transition metal oxides where a strong competition between FM and AFM interactions is observed. Long {\it et al}. suggested that DyCrO$_4$ undergoes two successive magnetic transitions: PM-FM transition at 23 K and  FM-AFM transition at 21 K (Ref. \cite{yl}). However, the neutron  diffraction results show that DyCrO$_4$ exhibits a single FM transition below 23 K with both the Cr$^{5+}$ and Dy$^{3+}$ spin moments  align collinearly along the $b$  axis and the magnetic interactions between Cr$^{5+}$ and Dy$^{3+}$  moments are mainly responsible for this FM transition \cite{ylon}. The large difference between the ZFC and FC magnetic susceptibility curves of DyCrO$_4$ may arise due to the domain wall pinning effect. The domains can align along the external field direction or in the local anisotropic field direction if the sample is cooled down from the high temperature with or without field which lead to irreversibility in susceptibility. Closer inspection reveals that there exists another very weak anomaly at around $T_{G}^{Dy}$$=$8 K. From ac susceptibility measurements, it has been suggested that a spin-glass like phase emerges below $T_{G}^{Dy}$  due to the strong competition between FM and AFM interactions \cite{yl}. The anomalies at $T_C^{Dy}$ and $T_{G}^{Dy}$ are more clearly reflected in $T$ dependence of d$\chi_{ZFC}/$d$T$ [inset of Fig. 1(a)].  For further understanding the nature of magnetic interactions, we have plotted $\chi^{-1}$ versus $T$ (Fig. 1). The linearity of $\chi^{-1}$ over a wide  range of $T$ above $T_C^{Dy}$ suggests that $\chi$ follows the Curie-Weiss (CW) law [$\chi$$=$C/($T-\theta_{CW}$)]. From the high temperature linear fit, we have calculated the CW temperature $\theta_{CW}$$\sim$3 K and the effective moment $\mu_{eff}$$=$10.7 $\mu$$_{B}$. $\chi$($T$)  for HoCrO$_4$ shows qualitative similar behavior, however, the ordering temperatures are smaller than that for DyCrO$_4$.  We find $T_C^{Ho}$$=$18 K and $T_{G}^{Ho}$$=$4 K. Likewise, for HoCrO$_4$, we have calculated  $\theta_{CW}$$\sim$0 K and  $\mu_{eff}$$=$10.5 $\mu$$_{B}$.  The observed values of $\mu_{eff}$ for both DyCrO$_4$ and HoCrO$_4$ are close to the expected moment (10.8 $\mu$$_{B}$) calculated from  the orbital moment contribution of the Dy$^{3+}$(4f) or Ho$^{3+}$(4f) ion and the spin only contribution of the Cr$^{5+}$(3d) ion with spin $S$$=$1/2. The small values of $\theta_{CW}$ indicate that  the strength of FM and AFM interactions in these compounds are comparable in magnitude. \\
\begin{figure}
\begin{center}
\includegraphics[width=.5\textwidth]{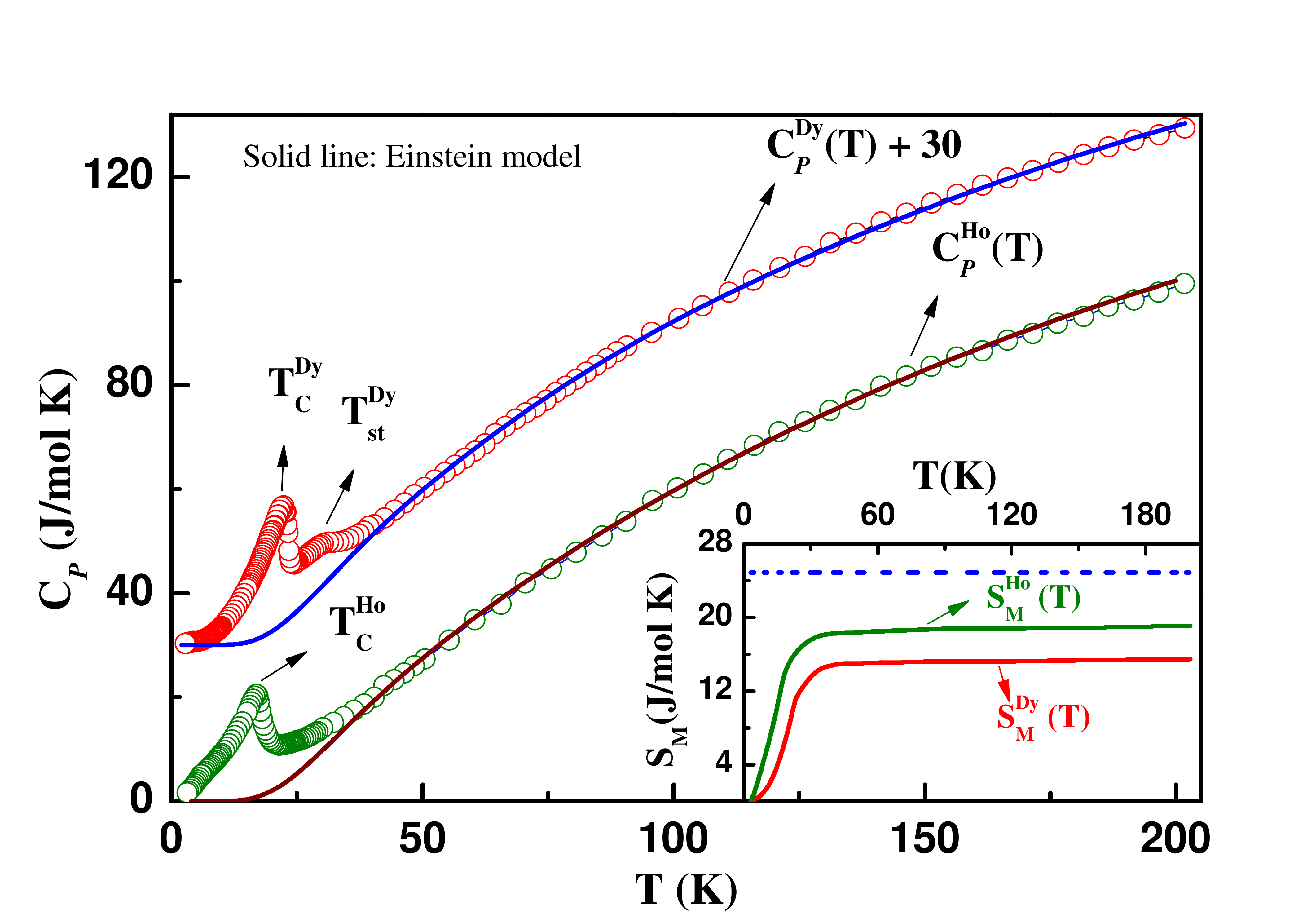}
\caption{ Zero-field heat capacity as a function of temperature for DyCrO$_4$ and  HoCrO$_4$. For clarity, we have shifted the heat capacity data for DyCrO$_4$ by 30 units along the $y$ axis. The solid lines  correspond to the fitting using the Einstein model for the lattice specific heat. Inset depicts the magnetic contribution of entropy  associated with the magnetic transition for both the compounds. The dotted line corresponds to the theoretical value of $S_M$.}
\end{center}
\end{figure}
Figure 2 describes the nature of temperature dependence of zero-field specific heat for DyCrO$_4$ and HoCrO$_4$ compounds. For DyCrO$_4$, $C_P$($T$) shows a $\lambda$-like peak close to  $T_C^{Dy}$. Another very weak anomaly is observed at $T_{st}^{Dy}$$=$31 K, which is consistence with the previously reported structural phase transition  from  tetragonal  to orthorhombic symmetry with decreasing temperature due to  the Jahn-Teller effect \cite{ylon,ktez}. $C_P$($T$) does not show any anomaly around 21 K due to AFM transition. This suggests that DyCrO$_4$ undergoes a single PM-FM transition just below 23 K.  For HoCrO$_4$, $C_P$($T$) also shows a $\lambda$-like peak around PM-FM transition, however, there is no indication of structural phase transition. $C_P$($T$) for both the compounds does not show any anomaly at low temperature due to the spin-glass like transition. For better understanding the nature of magnetic ground state,  the magnetic contribution to the specific heat ($C_M$)  in the vicinity of FM transition and beyond has been estimated. After subtracting the lattice contribution from $C_P$($T$), we obtain the value of $C_M$. The magnetic entropy ($S_M$) is obtained by integrating ($C_M/T)dT$. Inset of Fig. 2 shows that  $S_M$ tends to saturate at 60 and 75 $\%$ of theoretical value for DyCrO$_4$ and HoCrO$_4$, respectively. The overestimation of lattice specific heat may cause a reduction in magnetic entropy. However, this  is not significant. The main source of reduction in  entropy is the incomplete magnetic ordering near the transition. As these samples are prepared at a relatively low temperature, the grain size is expected to be small and the intergrain coupling is weak. For this reason, a significant contribution to the specific heat comes from the surface and intergranular regions which may have lower magnetic specific heat as compared to bulk and hence the reduction in entropy. The absence of saturation in $M$($H$) curve at 2 K up to $H$$=$ 8 T  and the smaller value of magnetic moment compared to the expected moment are also consistent with this picture.\\

As  the FM and AFM interactions in DyCrO$_4$ and HoCrO$_4$  are of comparable magnitude, one expects that the magnetic ground state of these compounds will be extremely sensitive to external perturbations such as magnetic field. In order to investigate the influence of magnetic field on magnetic ground state, we have measured $H$ dependence of $M$ in the neighborhood of the magnetic transition and beyond [Figs. 3 (a) and (b)]. For both the samples, at low temperatures well below $T_C$, $M$ increases rapidly above a critical field $H_c$ due to the metamagnetic transition from AFM to FM state. The approximate values of $H_c$  for DyCrO$_4$ and HoCrO$_4$ are 0.18 and 0.3 T, respectively at 2 K. Above $H_c$,  $M$($H$) exhibits a downward curvature and increases slowly without showing any saturation-like behavior up to the highest applied field and the value of magnetic moment decreases monotonically with increasing temperature. These are the characteristics of a ferromagnetic system.  At $T$$=$2 K and $H$$=$8 T, the  value of  $M$  is  8.0 $\mu_B$(8.3 $\mu_B$) per formula unit for DyCrO$_4$(HoCrO$_4$), which is smaller than the theoretical value (11 $\mu_B$). However, below $H_c$, $M$($H$) is weakly superlinear and $M$ does not decrease monotonically with $T$, which is in accord with the  field-induced transition from AFM to FM state at $H$$=$$H_c$.  The insets of Figs. 3 (a) and (b) show the five-segment  $M$($H$) loop at 2 K up to 2 T. $M$($H$) shows a small hysteresis at low fields, in particular, for HoCrO$_4$. It is observed that the hysteresis decreases rapidly with increasing temperature and disappears above 8 K. For this reason,  the $M$($H$) data during increasing field were plotted in Figs. 3 (a) and (b). To understand the nature of the field-induced magnetic transition, we have converted the $M$($H$) data into the Belov-Arrott plots  and the representative plots for  DyCrO$_4$ are shown in Fig. 3(c). The slope of the Belov-Arrott plot will be negative for a first-order magnetic phase transition, whereas it will be positive when the transition is second-order in nature. The positive slope of the Belov-Arrott plots at low  as well as high fields for both the samples implies that the field-induced FM transition is second-order in nature. \\
\begin{figure}
\begin{center}
\includegraphics[width=.45\textwidth]{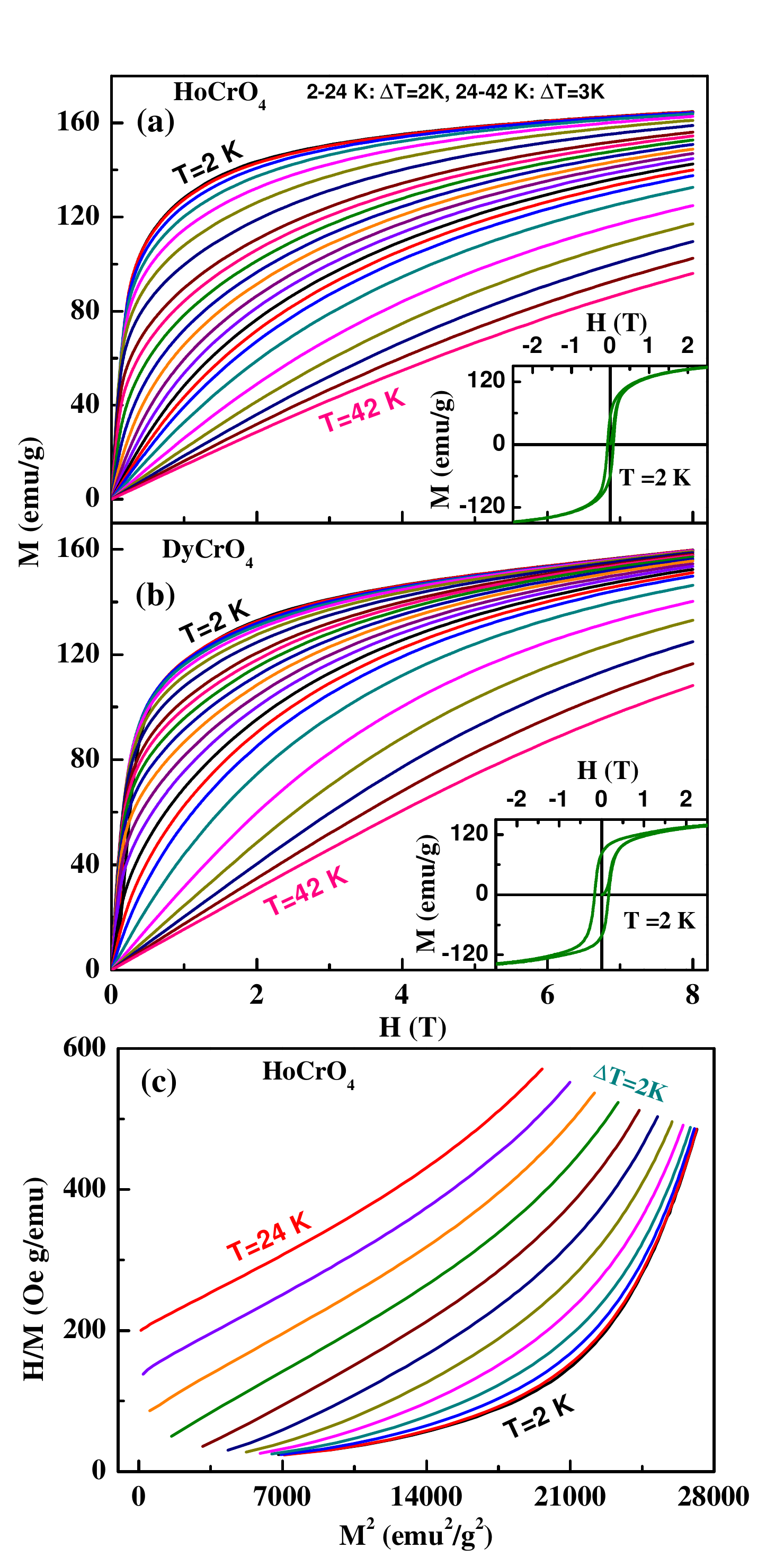}
\caption{ Isothermal magnetization for  DyCrO$_4$ (a) and  HoCrO$_4$ (b). Insets show the corresponding low-field hysteresis at 2 K. The Belov-Arrott plots for HoCrO$_4$ compound at some selected temperatures (c).}
\end{center}
\end{figure}
In order to test whether  DyCrO$_4$ and HoCrO$_4$ are suitable candidates for the magnetic refrigeration at liquid hydrogen temperature, we have calculated the isothermal magnetic entropy change as a function of temperature for field variations up to 8 T from the magnetization measurements [Figs. 3 (a) and (b)] using the well known relation,

\begin{equation}
\Delta S_{M}(T,H) = \sum_{i}\frac{M_{i+1} - M_{i}}{T_{i+1} - T_{i}}\Delta H_{i},
\label{eq2}
\end{equation}

where $M_{i+1}$ and $M_i$ are the experimentally measured values of magnetization with a field $H_{i}$ at temperatures $T_{i+1}$ and $T_{i}$, respectively. The thermal variation of  $\Delta S_{M}$ for  different magnetic field changes is shown in Figs. 4(a) and 4(b). For HoCrO$_4$, $\Delta S_{M}$ is negative down to the lowest measured temperature and the maximum value of $\Delta S_{M}$  ($\Delta S_{M}^{max}$) increases with field, reaching as high as 31.4 J kg$^{-1}$ K$^{-1}$  for a field change of 8 T.  For DyCrO$_4$, the qualitative nature of  $\Delta S_{M}$($T$) is similar to that for HoCrO$_4$, however, the value of $\Delta S_{M}^{max}$ is slightly smaller.  Also, in DyCrO$_4$, $\Delta S_{M}$ is small positive (inverse MCE) at low temperature.   Figure shows that the thermal distribution of $\Delta S_{M}$  for both the compounds is highly asymmetric with respect to the maximum.  With changing $T$, $\Delta S_{M}$ decreases at a  much faster rate below $T_C$ than above $T_C$. The steep decrease in $\Delta S_M$($T$) on the low-temperature side of the maximum and the small positive value of $\Delta S_M$  below 8 K for DyCrO$_4$ suggest that  AFM interaction is dominant at low temperature.  This kind of $T$ dependence of $\Delta S_M$ is observed in several antiferromagnetic systems where the field-induced AFM-FM transition occurs \cite{kag,midya1,jin,naik,anis}. When an external magnetic field is applied, the magnetic moment fluctuation is enhanced in one of the two AFM sublattices which is antiparallel to $H$. With  increasing $H$, more and more spins in the antiparallel sublattice orient along the field direction, which, in turn, increases the spin disordering and hence the inverse MCE occurs. For both the compounds, the maximum in $\Delta S_{M}$($T$) curve occurs at $T_C$ for low fields and it  shifts very slowly toward the higher temperature with increasing field.\\
\begin{figure}
\begin{center}
\includegraphics[width=.5\textwidth]{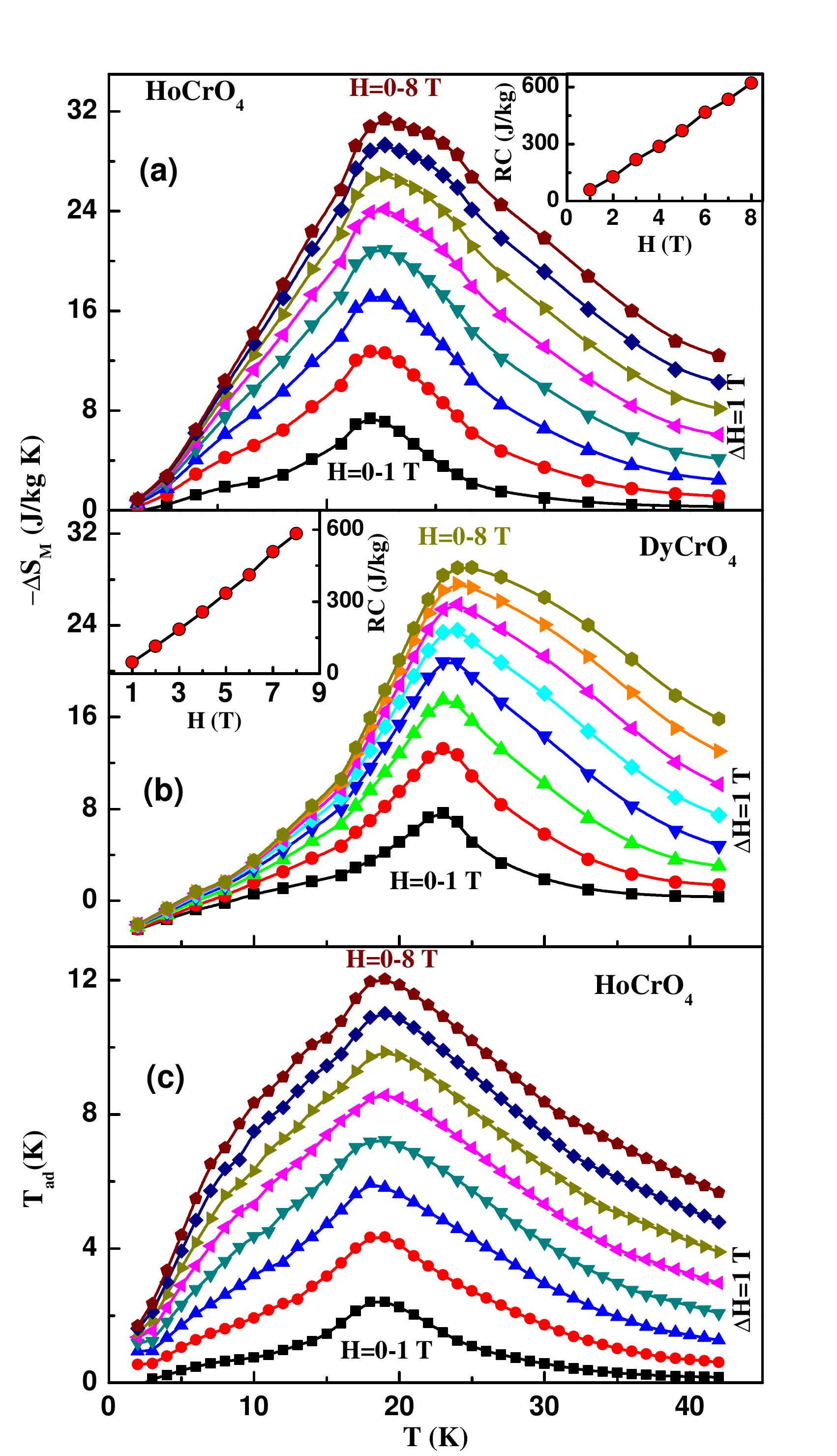}
\caption{ ITemperature dependence of isothermal magnetic entropy change $\Delta S_{M}$ for  HoCrO$_4$ (a) and  DyCrO$_4$ (b). Insets show the refrigerant capacity as a function of magnetic field.  Temperature dependence of adiabatic temperature change ($\Delta T_{ad}$) for HoCrO$_4$ (c).}
\end{center}
\end{figure}
The refrigerant capacity or relative cooling power is an important quality factor of the refrigerant material which determines the amount of heat transfer between the cold and hot reservoirs in an ideal refrigeration cycle and is defined as, RC$=\int\limits_{T_1}^{T_2}\Delta S_{M}dT$, where $T_1$ and $T_2$ are the temperatures corresponding to both sides of the half-maximum value of $\Delta S_{M}$($T$) peak. Insets of Figs. 4(a) and 4(b) show the magnetic field variation of refrigerant capacity of the materials. In both the cases, RC increases almost linearly with $H$.  The values of RC for a field change of 8 T are 622  and 583 J kg$^{-1}$ for HoCrO$_4$ and DyCrO$_4$, respectively.  In order to have better understanding on the application potential of $R$CrO$_4$ compounds, we have also calculated  MCE in terms of adiabatic temperature change $\Delta T_{ad}$ which is the isentropic temperature difference between $S(0,T)$ and $S(H,T)$. $\Delta T_{ad}$ may be calculated using the field-dependent magnetization and zero-field heat capacity data. $S(H,T)$ has been evaluated by subtracting the corresponding $\Delta S_M$($H$) from $S(0,T)$, where the total entropy $S(0,T)$ in absence of magnetic field is given by $S(0, T)=\int\limits_{0}^{T}\frac{C_p(0, T)}{T}dT$. The temperature dependence of $\Delta T_{ad}$ for HoCrO$_4$ is shown in Fig. 4(c) for various magnetic fields.  The maximum value of $\Delta T_{ad}$  ($\Delta T_{ad}^{max}$) reaches as high as 12 K for a field change of 8 T. For DyCrO$_4$, we observe that $\Delta T_{ad}^{max}$ is 8 K for a field change of 8 T. We may compare the present results on MCE with those reported for rare earth transition metal oxides with comparable $T_C$.  For example, in  multiferroic HoMnO$_3$ and DyMnO$_3$, the magnetocaloric parameters are quite large due to the field-induced AFM-FM transition of the rare earth sublattice but they are  significantly smaller than that for the present systems \cite{midya1,jin}. However, the  large values of magnetocaloric parameters for the present compounds are comparable with that for the magnetically frustrated  rare earth oxides EuHo$_2$O$_4$ and EuDy$_2$O$_4$  \cite{midya3}. In fact, under the same field change, the  values of $\Delta S_{M}^{max}$, RC, and $\Delta T_{ad}^{max}$ for the present compounds are much larger than that for most of the potential magnetic refrigerant materials with low magnetic transition temperature \cite{kag,shao,sbg,hz}.  It is also clear from Fig. 4 that the magnetocaloric parameters have reasonably large values at a moderate field strength which is an important criterion for magnetic refrigeration. We would like to mention that zircon-type $R$CrO$_4$ undergo pressure-induced structural phase transition to scheelite-type  and the value of field-induced magnetization for the scheelite-type phase is much larger than that for the zircon-type phase \cite{pasc}. So, the magnetic entropy change and hence, the MCE is expected to be significantly larger for the scheelite phase as compared to the zircon phase.\\

In conclusion,  we observe that the zircon-type DyCrO$_4$ and HoCrO$_4$ compounds exhibit complicated magnetic properties due to the strong competition between ferromagnetic and antiferromagnetic superexchange interactions of 3d and 4f spins. Both the compounds show field-induced metamagnetic transition at a relatively small applied field which leads to a giant negative entropy change near the magnetic transition. For HoCrO$_4$, the maximum  values of $\Delta S_M$, $\Delta T_{ad}$, and  RC are 31  J kg$^{-1}$ K$^{-1}$, 12 K, and 622 J kg$^{-1}$, respectively for a field change of 8 T. The magnetocaloric parameters are also quite large for a moderate field strength. The excellent magnetocaloric properties of $R$CrO$_4$  compounds can be utilized for the liquefaction of hydrogen.\\

The authors would like to thank S. Banerjee and G. N. Banerjee for their help during energy dispersive x-ray  study  and  thermogravity  analysis, respectively.


\begin{thebibliography}{99}
\bibitem{kag} K. A. Gschneidner Jr., V. K. Pecharsky, and A. O. Tsokol, Rep. Prog. Phys. {\bf 68}, 1479 (2005), and references therein.

\bibitem{tishin} A. M. Tishin, in Handbook of Magnetic Materials, edited by K. H. J. Buschow, (Elsevier Science B.V., New York, 1999), Vol. 12, p. 395

\bibitem{bfy} B. F. Yu, Q. Gao, X. Z. Meng, and Z. Chen, Int. J. Refrig. {\bf 68}, 622 (2003).

\bibitem{tk} T. Kimura, T. Goto, H. Shintani, K. Ishizaka, T. Arima, and Y. Tokura, Nature (London) {\bf 426}, 55 (2003).

\bibitem{ima} M. Imada, A. Fujimori, and Y. Tokura, Rev. Mod. Phys. {\bf 70}, 1039 (1998); M. B. Salamon and M. Jaime, {\it ibid}  {\bf 73}, 583 (2001).

\bibitem{lee} P. A. Lee, N. Nagaosa, and X. G. Wen, Rev. Mod. Phys. {\bf 78}, 17 (2006).

\bibitem{midya1} A. Midya, P. Mandal, S. Das, S. Banerjee, L. S. S. Chandra, V. Ganesan, and S. R. Barman, Appl. Phys. Lett. {\bf 96}, 142514 (2010).

\bibitem{jin} J. L. Jin, X. Q. Zhang, G. K. Li, Z. H. Cheng, L. Zheng, and Y. Lu, Phys. Rev. B {\bf 83}, 184431 (2011); A. Midya, P. Mandal, S. Das, S. Pandya, and V. Ganesan, {\it ibid} {\bf 84}, 235127 (2011); .

\bibitem{shao} M. Shao, S. Cao, S. Yuan, J. Shang, B. Kang, B. Lu, and J. Zhang, Appl. Phys. Lett. {\bf 100}, 222404 (2012).

\bibitem{ylon} Y. W. Long, Q. Huang, L. X. Yang, Y. Yu, Y. X. Lv, J. W. Lynn, Y. Chen, and C. Q. Jin, J. Magn. Magn. Mater.  {\bf 322}, 1912 (2010); M. Steiner, H. Dachs, and H. Ott, Solid State Commun.  {\bf 29}, 231 (1979).

\bibitem{pasc} E. C. Pascual, J. Romero de Paz, J. M. Gallardo Amores, and R. Sáez Puche, Solid State Sci.  {\bf 9}, 574 (2007).

\bibitem{yl} Y. Long, Q. Liu, Y. Lv, R. Yu, and  C. Jin, Phys. Rev. B  {\bf 83}, 024416 (2011).

\bibitem{ktez} K. Tezuka and Y. Hinatsu, J. Solid State Chem.  {\bf 160}, 362 (2001).

\bibitem{naik} V. B. Naik, S. K. Barik, R. Mahendiran, and B. Raveau, Appl. Phys. Lett. {\bf 98}, 112506 (2011).

\bibitem{anis} A. Biswas, S. Chandra, T. Samanta, B. Ghosh, S. Datta, M. H. Phan, A. K. Raychaudhuri, I. Das, and H. Srikanth,
Phys. Rev. B. {\bf 87}, 134420 (2013).

\bibitem{midya3} A. Midya, N. Khan, D. Bhoi, and P. Mandal, Appl. Phys. Lett. {\bf 101}, 132415  (2012).

\bibitem{sbg} S. B. Gupta, and K. G. Suresh, Appl. Phys. Lett. {\bf 102}, 022408 (2013).

\bibitem{hz} H. Zhang, B. G. Shen, Z. Y. Xu, J. Shen, F. X. Hu, J. R. Sun, and Y. Long, Appl. Phys. Lett. {\bf 102}, 092401 (2013).

\end{thebibliography}
\end{document}